\documentclass[final,3p,times,twocolumn]{elsarticle}

\usepackage{amssymb}
\usepackage{amsmath}
\usepackage{amsthm}
\usepackage{float}

\usepackage{lineno}
\usepackage{xcolor}

\journal{Science Bulletin}

\begin{document}

\begin{frontmatter}



\title{A Novel Quantum Realization of Jet Clustering in High-Energy Physics Experiments}

\author[inst1]{Yongfeng Zhu}

\affiliation[inst1]{organization={State Key Laboratory of Nuclear Physics and Technology, School of Physics, Peking University},
            city={Beijing 100871},
            country={China}}

\author[inst2]{Weifeng Zhuang}
\author[inst2]{Chen Qian}
\author[inst2]{Yunheng Ma}
\author[inst2,inst3]{Dong E. Liu\corref{cor1}}
\cortext[cor1]{Corresponding authors}
\ead{dongeliu@mail.tsinghua.edu.cn}
\author[inst4,inst5]{Manqi Ruan\corref{cor1}}
\ead{ruanmq@ihep.ac.cn}
\author[inst1]{Chen Zhou\corref{cor1}}
\ead{czhouphy@pku.edu.cn}

\affiliation[inst2]{organization={Beijing Academy of Quantum Information Sciences},
            city={Beijing 100193},
            country={China}}

\affiliation[inst3]{organization={Department of Physics, Tsinghua University},
            city={Beijing 100084},
            country={China}}

\affiliation[inst4]{organization={Institute of High Energy Physics, Chinese Academy of Sciences},
            city={Beijing 100049},
            country={China}}

\affiliation[inst5]{organization={University of Chinese Academy of Sciences},
            city={Beijing 100049},
            country={China}}

\end{frontmatter}



\begin{figure*}[htbp]
    \centering
    \includegraphics[width=.9\textwidth]{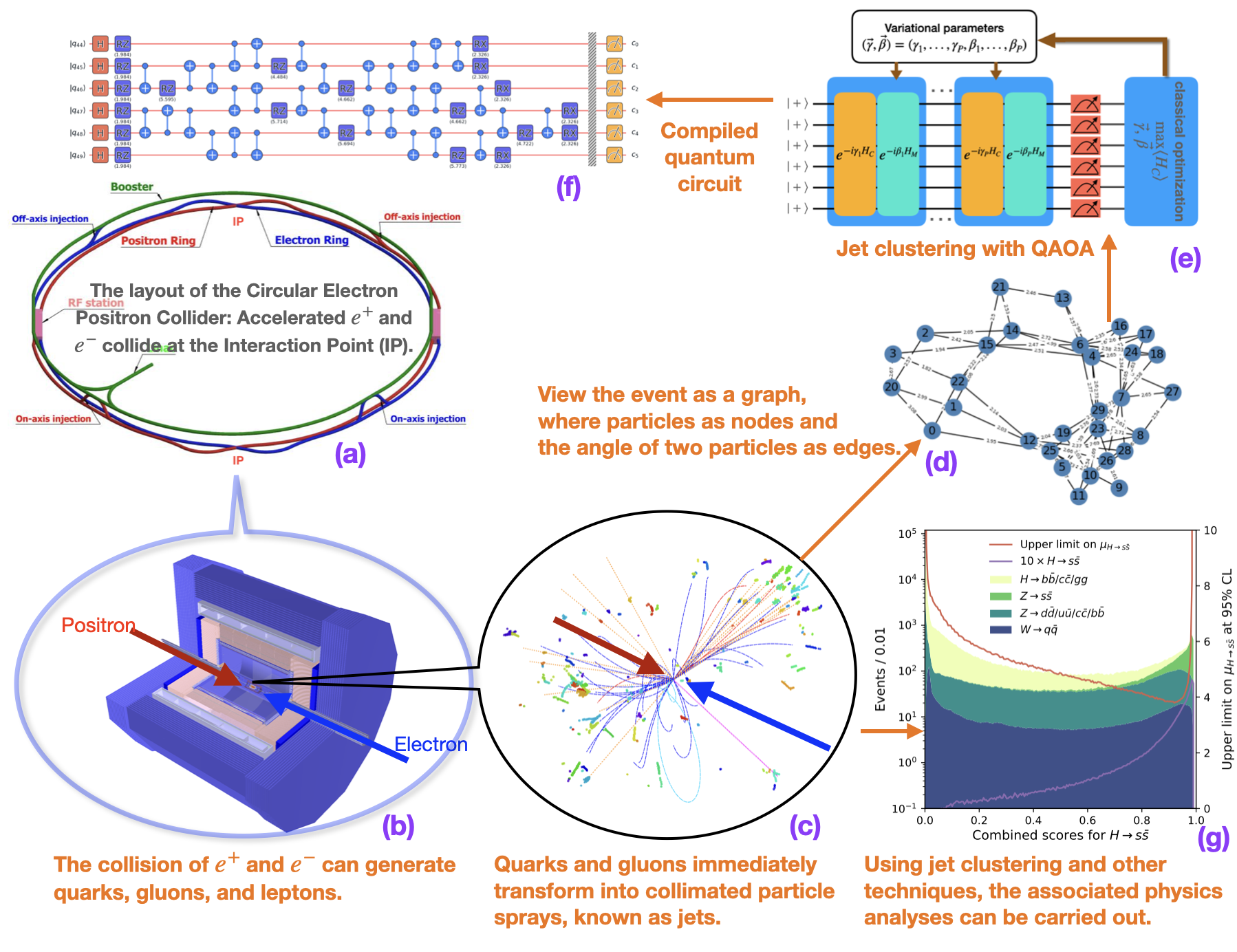}
    \caption{
    A typical physics analysis procedure involving jets in high-energy physics is outlined:
    Panel (a) \cite{CEPCStudyGroup:2018ghi} depicts the layout of the Circular Electron-Positron Collider. Electrons and positrons are accelerated in the booster and collide at the Interaction Point (IP) within the detector, illustrated in panel (b) \cite{CEPCStudyGroup:2018ghi}. Quarks and gluons resulting from these collisions transform into collimated streams of particles known as jets. Panel (c) illustrates an $e^+e^-\to ZH(Z\to \nu\bar{\nu}, H\to s\bar{s})$ event, where the two quarks form two jets. Jets are clustered using an algorithm that groups final-state particles into sets corresponding to individual gluons or quarks. During jet clustering, each event is represented as a graph, as shown in panel (d). The Quantum Approximate Optimization Algorithm (QAOA) can be applied to perform jet clustering, as demonstrated in panel (e). The compiled quantum circuit is shown in panel (f). The information from reconstructed jets can be utilized to achieve related objectives, such as the $H\to s\bar{s}$ measurement shown in panel (g) \cite{PhysRevLett.132.221802}. 
\label{fig:process}}
\end{figure*}

Application of quantum technologies to fundamental sciences has the potential to advance both fields simultaneously.
In the near future, the number of reliable quantum operations in a real-world quantum computer will be constrained by noise and decoherence \cite{Krantz2019AQE}.
To overcome these challenges, hybrid quantum-classical algorithms \cite{McClean:2015vup} have been proposed.
A notable example is the Quantum Approximate Optimization Algorithm (QAOA) \cite{Farhi:2014ych}, an algorithm commonly used to address classical combinatorial optimization problems. 
In high-energy particle collisions, quarks and gluons are produced and immediately form collimated particle sprays, referred to as jets, due to color confinement \cite{PhysRevD.10.2445}. 
Accurate jet clustering is crucial to reconstruct the quark or gluon information and enhances the study of the properties of the Higgs boson. 
For the first time, we apply QAOA to the problem of jet clustering, as shown in Fig.~\ref{fig:process}e, by mapping collision events into graphs, as shown in Fig.~\ref{fig:process}d. 
We obtain experimental results on quantum computer simulators and quantum computer hardware and find that their performance is comparable to or even better than classical algorithms for a small-sized problem. 

\renewcommand{\dblfloatpagefraction}{.9}

A typical physics analysis procedure involving jets in high-energy physics is as follows.
Data is collected from an electron-positron Higgs factory, such as the Circular Electron-Positron Collider (CEPC) \cite{CEPCStudyGroup:2018ghi}, whose layout is shown in Fig.~\ref{fig:process}a.
Electrons and positrons are accelerated in the booster and interact at the Interaction Point (IP), which is located within the detector, as illustrated in Fig.~\ref{fig:process}b. 
The golden process for studying Higgs properties is $e^+e^-\to ZH$.
An $e^+e^-\to ZH(Z\to \nu\bar{\nu}, H\to s\bar{s})$ event, where two quarks form two jets, is displayed in Fig.~\ref{fig:process}c. 
A jet clustering algorithm, which groups final-state particles into sets corresponding to individual gluons or quarks, is applied to identify these jets. 
Following jet clustering, various physics analyses can be performed, such as the measurement of the branching ratio of $H\to s\bar{s}$, shown in Fig.~\ref{fig:process}g.
Given the continuous progress in quantum computing technology, there is increasing interest in identifying practical applications for near-term quantum machines.
For instance, the calculation of thrust, a variable was formulated both as a quantum annealing problem and a Grover search problem \cite{PhysRevD.101.094015}.
Other studies \cite{Delgado:2022snu,Pires:2020urc} have developed quantum annealing-based jet clustering algorithms.

The QAOA draws inspiration from the Quantum Adiabatic Algorithm (QAA).
QAA uses two Hamiltonians: an initial Hamiltonian $\hat{H}_M$ with an easily prepared ground state, and a final Hamiltonian $\hat{H}_C$ whose ground state encodes the solution to the optimization problem.
The adiabatic evolution path is defined by the transitional Hamiltonian $\hat{H}(t) = s(t)\hat{H}_C + (1-s(t))\hat{H}_M$, where $s(t) = \frac{t}{T}$, $t\in [0, T]$, and T is the total evolution time.
The evolution operator $\hat{U}(t)$ is derived from $\hat{H}(t)$.
QAA can be emulated on a gate-based quantum computer by defining $\hat{U}(t)$ into sufficiently small steps,
\begin{equation}
\label{eq:evolution}
\begin{split}
\hat{U}(t) & \approx \prod_{j=1}^{P} e^{-i\hat{H}(j\Delta t)\Delta t}  = \prod_{j=1}^P e^{-i(1-s(j\Delta t))\hat{H}_M\Delta t}e^{-is(j\Delta t)\hat{H}_C\Delta t} \\ & =\prod_{j=1}^P e^{-i\beta_j\hat{H}_M}e^{-i\gamma_j\hat{H}_C} = \prod_{j=1}^P \hat{U}_M(\beta_j) \hat{U}_C(\gamma_j)
 \end{split}
\end{equation}
where $\Delta t = T/P$ and $P$ is the number of steps.
QAOA was designed from this trotterized version with repeated cost $\hat{U}_C(\gamma)$ and mixer layers $\hat{U}_M(\beta)$ \cite{Farhi:2014ych}.

An undirected graph $G=(V, E)$, where $V$ is the set of nodes and $E$ is the set of edges, can encode a collision event in high-energy physics.
The nodes represent particles and the edges represent pairs of particles that are nearby.
To define the edges, a weight $w_{ij}$ is assigned to each pair of particles that is equal to the angle between them.
Then, for each node, edges are sorted by weight in decreasing order and only the top k edges are retained (k-regular graph).
Figure~\ref{fig:process}e illustrates an $e^+e^-\to ZH(Z\to \nu\bar{\nu}, H\to s\bar{s})$ event with 30 particles and k=3.
After mapping collision events into graphs, QAOA can perform jet clustering akin to solving the Max-{\it K}-cut problem \cite{Proietti:2022kpj}.
This study clusters two jets, analogous to the Max-cut problem (the specific case of the Max-{\it K}-cut problem with {\it K} equal to 2).
In other words, the algorithm aims to partition the graph nodes into two complementary subsets that maximize the weighted sum of edges connecting nodes in different subsets, given by 
$C(x)=\sum_{i,j=1}^{|v|}w_{ij}x_i(1-x_j)$.
Maximizing $C(x)$ is NP-hard, but QAOA can approximate $C(x)$ close to its maximum value $C_{max}=\mathop{\text{max}}\limits_xC(x)$~\cite{Farhi:2014ych}.
To solve the Max-cut problem using QAOA, we begin by defining the problem Hamiltonian as $\hat{H}_C = \frac{1}{2}\sum_{(i,j)\in E}W_{ij}(I-\sigma^z_i\sigma^z_j)$ and initial Hamiltonian as $\hat{H}_M = \sum_{j=1}^n\sigma_j^x$, where $\sigma^z_i$ represents the Pauli-Z operator acting on the i-th qubit and $W_{ij}$ represents the angle between particles i and j. 
By substituting $\hat{H}_C$ and $\hat{H}_M$ into the evolution operator shown in Equation \ref{eq:evolution} and decomposing it into sequences of quantum gate operations, we construct the corresponding quantum circuit.
This circuit is then used to evolve the system's initial state toward the ground state of $\hat{H}_C$, with a classical optimizer iteratively adjusting the parameters $\gamma$ and $\beta$.
For practical jet clustering, the focus is on approximating the ground state.
Among the sampling states, the one closest to the ground state is selected.
The procedure is illustrated in Fig.~\ref{fig:process}e and more details are in~\cite{Farhi:2014ych,Fuchs2020EfficientEO}.
The key parameters are the number of trotterized steps (QAOA depth), the evolution times (optimized by a classical optimization algorithm COBYLA~\cite{Bonet-Monroig:2021yfd} and an interpolation method~\cite{Fuchs2020EfficientEO}), and the number of times sampling is performed (1024 in this study).
Jet clustering performance is evaluated based on the angles between the reconstructed jets and the corresponding quarks.
In the process $e^+e^-\to ZH(Z\to \nu\bar{\nu}, H\to s\bar{s})$, each event produces two jets, resulting in two angles $angle_1$ and $angle_2$.
This study uses the sum $= angle_1 + angle_2$ as the performance metric.

\begin{figure*}[htbp]
    \centering
    \includegraphics[width=0.9\textwidth]{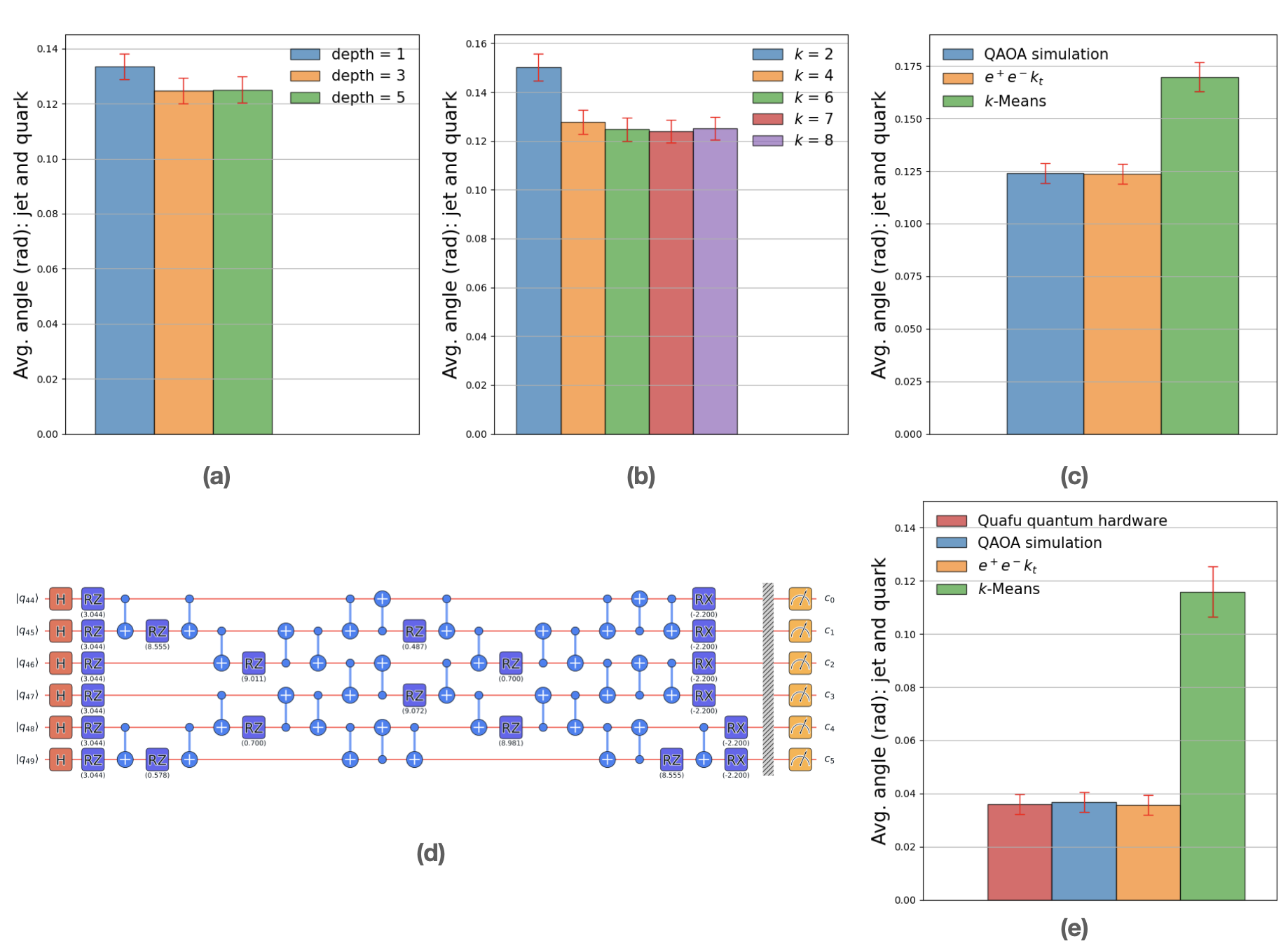}
    \caption{ Panels (a), (b), and (c) show the jet clustering performance for 4000 events with 30 particles. (a) The jet clustering performance with k set to 6 and QAOA depths of 1, 3, and 5. (b) The performance with QAOA depth set to 3 and k set to 2, 4, 6, 7, and 8. (c) Comparison of the QAOA (with depth = 5 and k = 7), the $e^+e^-k_t$ algorithm, and the k-Means algorithm. (d) Compiled quantum circuit for performing QAOA-based jet clustering on a 6-particle event, based on the {\it Baihua} quantum hardware. The circuit has 34 CNOT gates, 27 single-qubit gates, and a depth of 27. (e) Jet clustering performance for 1217 6-particle events with the QAOA (with depth = 1 and k = 2) run on a quantum computer and a quantum simulator, and with the classical algorithms. \label{fig:ensemble}}
\end{figure*}

The QAOA performance can be enhanced by increasing depth and decreasing the evolution times, allowing for more granular unitary totter increments.
This granularity aids in retaining the quantum system's state within the vicinity of the ground state. 
Figure~\ref{fig:ensemble}a compares jet clustering performance on depths of 1, 3, and 5 with k = 6.
The results for depths 3 and 5 are comparable and noticeably superior to that of depth 1. 

A higher value of k signifies that a node is linked to more nodes within the graph, indicating a more complex quantum circuit. 
Figure~\ref{fig:ensemble}b compares the performance of the jet clustering for k of 2, 4, 6, 7, and 8 with the depth fixed at 3.
The optimal jet clustering performance is at k of 7.

Figure~\ref{fig:ensemble}c displays the performance of QAOA with depth = 3 and k = 7 together with that of two classical algorithms: k-Means and $e^+e^-k_t$ \cite{Cacciari:2011ma}. 
The k-Means algorithm partitions a dataset into k clusters by minimizing the sum of squared distances between data points and the centroids of their respective clusters.
For our jet clustering study, the angle between two particles is used as the distance metric.
The $e^+e^-k_t$ algorithm identifies the pair of closest particles, combines them, and then repeats this procedure until some stopping criterion is reached.
 The distance metric in this algorithm is $d_{ij} = 2min(E_i^2, E_j^2)(1-cos\theta_{ij})$, where $E_{i/j}$ represents the energy of particle i/j and $\theta_{ij}$ represents the angle between particles i and j.
From Fig.~\ref{fig:ensemble}c, the performance of QAOA is comparable to that of the $e^+e^-k_t$ algorithm and better than that of the k-Means algorithm.
This comparison highlights the potential of the QAOA in the jet clustering problem.

Finally, we conduct the analysis on the {\it BAQIS Quafu} quantum computing cloud, which provides users access to superconducting quantum processors.
The {\it Baihua} processor features 123 operational qubits interconnected through couplers.
The relaxation time $T_1$ and the dephasing time $T_2^\ast $ characterize the decoherence of the qubit state.
This chip has an average $T_1$ of 73.994 microseconds and an average $T_2^\ast$ of 29.02 microseconds.
The single-qubit gates have an average fidelity (probability of correct output) of 99.9\% and the two-qubit gates (CZ) have an average fidelity of 98.8\%.
A quantum circuit implementing QAOA was compiled with the {\it Quafu-Qcover} compiler \cite{Hong-Ze:50302}. 
The compiled quantum circuit is then transformed into a series of qubit operation commands and sent to the quantum control system of the quantum computer.
The quantum control system executes these commands sequentially, with each command corresponding to a pulse operation used to control the qubits.

An ideal quantum processor, free from noise and with sufficient coherence time, can utilize all available qubits for computations. 
However, due to current hardware limitations such as noise and limited connectivity--where each qubit can only interact with its immediate 2 or 3 neighbors--we restricted our study to 6 qubits.
For events involving 6 particles, we found that the QAOA performance is already optimal when the QAOA depth is set to 1 and k to 2.
Figure~\ref{fig:ensemble}d illustrates a compiled quantum circuit of QAOA for performing jet clustering on a 6-particle event.
The compiled circuit has 34 CNOT gates, 27 single-qubit gates, and a depth of 26.
Figure~\ref{fig:ensemble}e showcases the performance of this circuit,  run on the {\it Baihua} processor, applied to 1217 6-particle events.
For this small-sized problem, the quantum hardware achieves similar performance to a noiseless quantum computer simulator.

In summary, the rapid development of quantum algorithms and hardware devices enables quantum computers to solve small-scale but representative problems in fundamental science.
For example, the QCSH package \cite{Lv2023} provides an efficient quantum algorithm for calculating nuclear structure. 
A quantum algorithm for heat conduction \cite{Wei2023AQA} significantly outperforms classical counterparts.
An algorithm for simulating ocean circulation on a quantum computer \cite{ocean} and a quantum circuit for implementing spectral clustering \cite{specClus} both achieve significant speedup over classical counterparts.
We report the first application of QAOA to the problem of jet clustering.
We found that QAOA performance based on a simulation of a quantum computer with 30 qubits is close to the $e^+e^-k_t$ algorithm and superior to the k-Means algorithm.
The QAOA was also tested on hardware using 6 qubits on the {\it BAQIS Quafu} quantum computing cloud.
The performance of the algorithm on the quantum hardware is similar to the performance on a noiseless quantum simulator for a small-sized problem.
As quantum hardware and algorithms continue to advance, we expect to be able to tackle larger jet clustering problems.
This study marks a significant step toward a practical quantum computing application in high-energy physics. 

\section*{Conflict of interest}

The authors declare that they have no conflict of interest.

\section*{Acknowledgment}
This work is supported by the Fundamental Research Funds for Central Universities, Peking University, the state Key Laboratory of Nuclear Physics and Technology, Peking University (Grant No. NPT2022ZZ05), the National Key R\&D Program of China under Contracts No. 2022YFE0116900, the National Natural Science Foundation of China (NSFC) under Grant No. 12305010, the Key Laboratory for Particle Astrophysics and Cosmology (Ministry of Education), Shanghai Key Laboratory for Particle Physics and Cosmology, and the Beijing Natural Science Foundation under Grant No. Z220002.

We thank Hongze Xu for his invaluable support in utilizing Quafu Cloud and Andrew Levin for his meticulous efforts in refining the grammar and expressions in our manuscript.
We extend our appreciation to the Beijing Academy of Quantum Information Sciences for access to quantum hardware and to the Computing Center at the Institute of High-Energy Physics for providing computing resources.

\section*{Author contributions}
Yongfeng Zhu analyzed the data and wrote the initial manuscript. Weifeng Zhuang,
Chen Qian and Yunheng Ma researched QAOA and discussed it with Yongfeng Zhu. Dong E. Liu, Manqi Ruan, and Chen Zhou conceived of and oversaw the project. All authors contributed to analyzing the results and to revising and reviewing the manuscript.



\end{document}